%% file: mars2017.tex
\documentclass[submission,copyright,creativecommons]{eptcs}
\usepackage{prelude}
\providecommand{\event}{MARS 2017} 

\begin{document}
\input{title}
\input{intro}
\input{framework}
\input{actions}
\input{extensibility}
\input{experiments}
\input{concl}
\nocite{*}
\bibliographystyle{eptcs}
\bibliography{mars2017References}
\end{document}

%% file: title.tex
\title{A Model-Derivation Framework for Software Analysis}

\author{
	Bugra M.Yildiz\textsuperscript{1} \qquad Arend Rensink\textsuperscript{1} \qquad Christoph Bockisch\textsuperscript{2} \qquad Mehmet Aksit\textsuperscript{1}
	\institute{\textsuperscript{1}Formal Methods and Tools Group, University of Twente}
	\institute{\textsuperscript{2}Faculty of Mathematics and Computer Science, Philipps-Universit{\"a}t Marburg}
	\email{\{b.m.yildiz, arend.rensink, m.aksit\}@utwente.nl}
	\email{bockisch@mathematik.uni-marburg.de}
}	
\def\authorrunning{B. M. Yildiz, A. Rensink, C. Bockisch, M. Aksit}
\def\titlerunning{A Model-Derivation Framework for Software Analysis}
\maketitle

\begin{abstract}
Model-based verification allows to express behavioral correctness conditions like the validity of  execution states, boundaries of variables or timing at a high level of abstraction and affirm that they are satisfied by a software system. However, this requires expressive models which are difficult and cumbersome to create and maintain by hand. This paper presents a framework that automatically derives behavioral models from real-sized Java programs. Our framework builds on the EMF/ECore technology and provides a tool that creates an initial model from Java bytecode, as well as a series of transformations that simplify the model and eventually output a timed-automata model that can be processed by a model checker such as \uppaal. The framework has the following properties: (1) consistency of models with software, (2) extensibility of the model derivation process, (3) scalability and (4) expressiveness of models. We report several case studies to validate how our framework satisfies these properties.

\end{abstract}

%% file: intro.tex
\section{Introduction}

One of the main challenges in developing a software system is to ensure that it fulfills the specifications. Validation of software systems by testing is generally considered to be a labor-intensive and tedious task \cite{Ciortea2010}. For this reason, model-based verification techniques have been introduced, which aim at verifying software systems through the use of models, instead of testing at the implementation level \cite{utting2010practical}. Such approaches naturally require the existence of expressive models of the systems being considered.

Unfortunately, deriving expressive models for software systems  for the purpose of verification is not a trivial task \cite{sargent2011}. Firstly, models are typically defined through a manual effort. The modeler must be an expert in the adopted modeling technique, must have a deep understanding of the software being modeled and must have skills for abstracting away the unnecessary details. These challenges make the model building process a labor-intensive and error-prone task. As a result, models of the same system can vary depending on the skills and preferences of the modeler. Secondly, software systems evolve continuously. Models must be maintained in parallel or else they become outdated \cite{rech2011}. Keeping models consistent with software is tedious work.

To counter these challenges, we propose the use of automatic model derivation from program code. This paper presents a framework that automatically derives models from Java programs that can be used by a model checker. There are a number of similar proposals \cite{bernat, tetaJ, bucaioni3952, corbett2000} that share the overall aim of our framework. However, no evidence is reported that they are capable of handling large programs. We show that our framework can, for instance, derive a model of a Java program with around 1500 classes in a reasonable time. (Note that the actual analysis of the generated models is not the focus of this paper.)

We have built the implementation of this framework upon the MDE technology called Epsilon/EMF \cite{epsilonWebsite}. As part of our framework, we have developed an ECore metamodel of Java bytecode. As target formalism for the derived models, we have chosen timed automata, since we want to conduct (among other things) timing analysis in future work of our project \cite{tips}. Our framework produces models compatible with the \uppaal model checker \cite{uppaalWebsiteRef}.

\paragraph{Overview.}

The elements of the framework are shown in Figure \ref{figTifFlow}. The actions, described in some more detail in Section~\ref{sectionConsistency}, are as follows: 

\begin{figure}
	\centering
	\includegraphics[height=5.5cm]{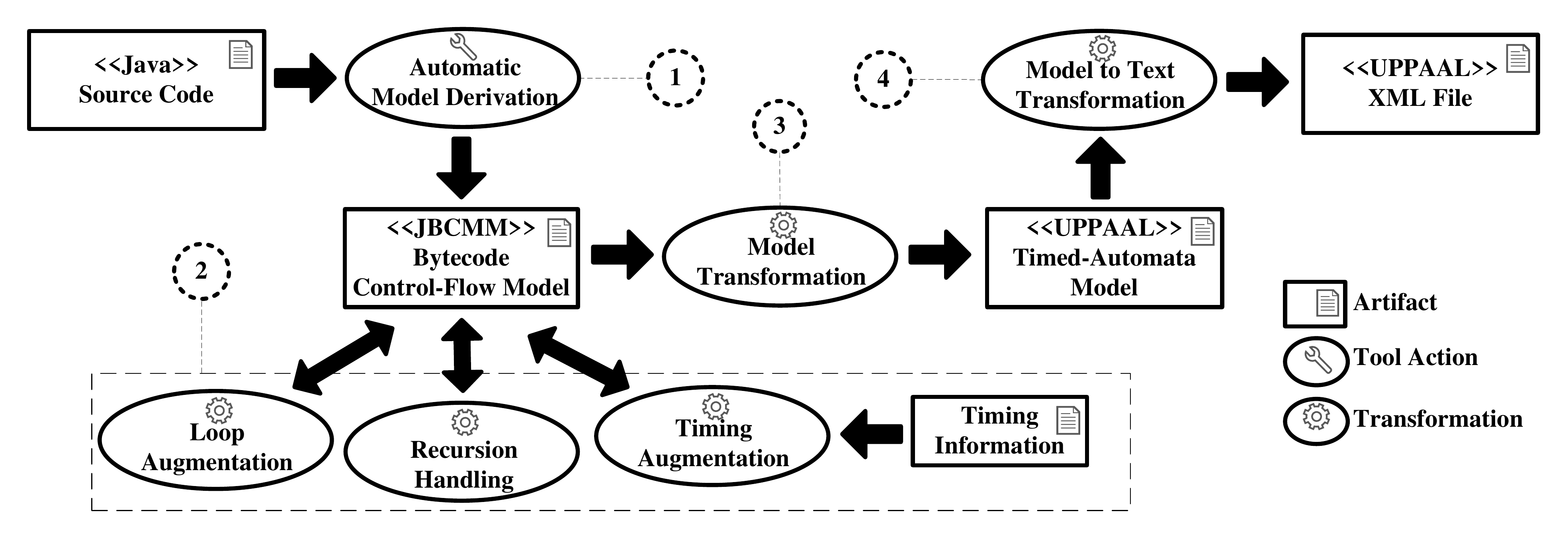}
	\caption{The automatic model-derivation framework}
	\label{figTifFlow}
\end{figure}

\begin{itemize}
\item Action~1 automatically derives a bytecode control-flow model of the software system from its bytecode.
\item Action 2 modifies and enriches the bytecode control-flow model with properties serving, e.g., as a prerequisite for the model-based timing analysis: \begin{enumerate*}[label=\itshape\alph*\upshape)] \item loop augmentation, to detect the loops in the control-flow model and annotate them with repetition limits; \item recursion handling, to modify the control-flow model for feasible model checking by detecting the recursive calls via a derived call graph and handling them, \item timing augmentation inserts the predictions of execution times into the control-flow model\end{enumerate*}.
\item Action 3 transforms the control-flow model to a timed-automata model. This is shown on an example in Section~\ref{sectionModelDerivation}.
\item Action 4 transforms this model to the input format of the model checker.
\end{itemize}

\paragraph{Contribution and roadmap.}

We claim the following benefits for our model derivation framework:

\begin{enumerate}[itemsep=2pt]
\item \emph{Consistency}: Our framework ensures that the derived models are consistent with the original software. Since all the individual actions of the framework are fully automated, it is easy to re-create models after changes occur in the code. We elaborate on this in Section~\ref{sectionConsistency}.

\item \emph{Extensibility}: Our framework can be extended in a systematic way to adapt to various analysis needs. We discuss this further in Section~\ref{sectionExtensibility}, by providing some example extensions.

\item \emph{Scalability}: Our framework can handle large Java programs in an acceptable time span. The time needed to check a model is known to grow exponentially with the size of the model. For this reason, Action~2 of our framework provides model transformations that simplify the control-flow model before the \uppaal model is derived in Action~3. Furthermore, we generate timed-automata models such that optimizations applied by \uppaal are applicable. We elaborate this in Section \ref{sectionScalability}.

\item \emph{Expressiveness}: The derived models are expressive enough for analysis purposes. Essentially, bytecode instructions are transformed into locations, in such a way that \uppaal queries can be formulated about the timing properties of the model that reflect real properties of the original system. We discuss this further in Section \ref{sectionExpressiveness}.
\end{enumerate}
For more detailed information, please refer to our technical report \cite{eemcs26622}. The link to the repository of the framework is available at \cite{ourRepository}.

\paragraph{Related work.}

There are a number of automatic model derivation tools for analysis of software systems \cite{bernat, tetaJ, bucaioni3952, corbett2000}, which share the overall aim of our framework. However, no evidence is reported for these tools showing that they are capable of handling large programs.

Corbett et al. introduced Bandera \cite{corbett2000}, an integrated collection of analysis components for Java programs. Bandera produces finite-state models in the input language of several verification tools from the source code of Java programs. One major difference with our framework is that timing analysis is not aimed by Bandera through its generated models. Another difference is that Bandera derives its models from the source code but not from the bytecode of Java programs. As a result of this, it does not offer to extend its analysis to include the third-party components whose source code is not available or it does not analyze other non-Java language programs which compile to bytecode. Lastly, the study reports the timing performance of model-derivation for a small example, however there is no information reported on the scalability of the model-derivation process for large code sizes.

Frost et al. presented the tool called TetaJ \cite{tetaJ} for static analysis of Java programs using model checking and Luckow et al. presented the tool called TetaSARTS in \cite{tetaSARTS} as a continuation of TetaJ to address schedulability of Java programs. They use three layers of models that are presented as separate templates in a timed-automata model in \uppaal: the program, the virtual machine and the hardware. The program model is automatically derived from the bytecode and the mapping of bytecode elements to the timed-automata model is done in a similar approach as we do. Although their tool includes some optimizations coming with it to reduce the state-space size, it does not offer mechanisms for extensions. They report the performance results of the model checking some examples up to 18 classes and 44 methods, but the papers include no information about the performance and scalability of the automatic model derivation process for large programs. 

Bucaioni provides a tool developed using MDE techniques to generate models of a vehicular embedded application in order to perform timing analysis \cite{bucaioni3952}. However, neither are there an explicit extension mechanisms offered nor do the authors report about the scalability of timing performance.

Bernat et al. proposed a WCET analysis scheme based on Java bytecode and a tool called Javelin to support this scheme \cite{bernat}. They use static method calls to inject the missing data information in the bytecode such as loop iterations. They focus on WCET analysis only and do not use model checking for this purpose. Javelin does not offer any extension mechanisms. There is also no explanation on how inheritance and polymorphism are handled and nor do they report on the scalability of their approach.

%% file: framework.tex
\section{Model Derivation Framework}\label{sectionModelDerivation}

The core transformation of our framework is Action~3, which derives an \uppaal model from an (enriched) bytecode control-flow model. We explain the basics of this transformation very briefly; for more information please refer to \cite{eemcs26622}.

A timed-automata model in \uppaal consists of synchronized instances of templates, where each template defines a single timed automaton; the synchronization is based on transition labels. 

\begin{itemize}

\item The \uppaal model derived from a Java program consists of a template for every Java class, plus one global template to maintain a global clock and kick off the program. 

\item Each class-derived template consists of a loop for each method, starting with an action corresponding to a call of that method, expressed as \snip{<class_name>\hash <method_signature>\hash call?} and ending with the corresponding \snip{<class_name>\hash <method_signature>\hash return!}. 

\item The locations correspond to nodes of the bytecode control flow graph and are labeled \snip{l_<line_number>_<index>}; the transitions correspond to flow graph edges. 

\item If a location corresponds to a method call, each incoming transition is labeled with the call actions, and each outgoing transition with the return actions.

\item A local clock per template, \snip{lc}, is used in invariant and guard expressions to force the system to spend between \snip{tlb_<location_name>} and \snip{tub_<location_name>} time units at a particular location. 

\end{itemize}

\begin{figure}
\centering
\includegraphics[height=5.7cm]{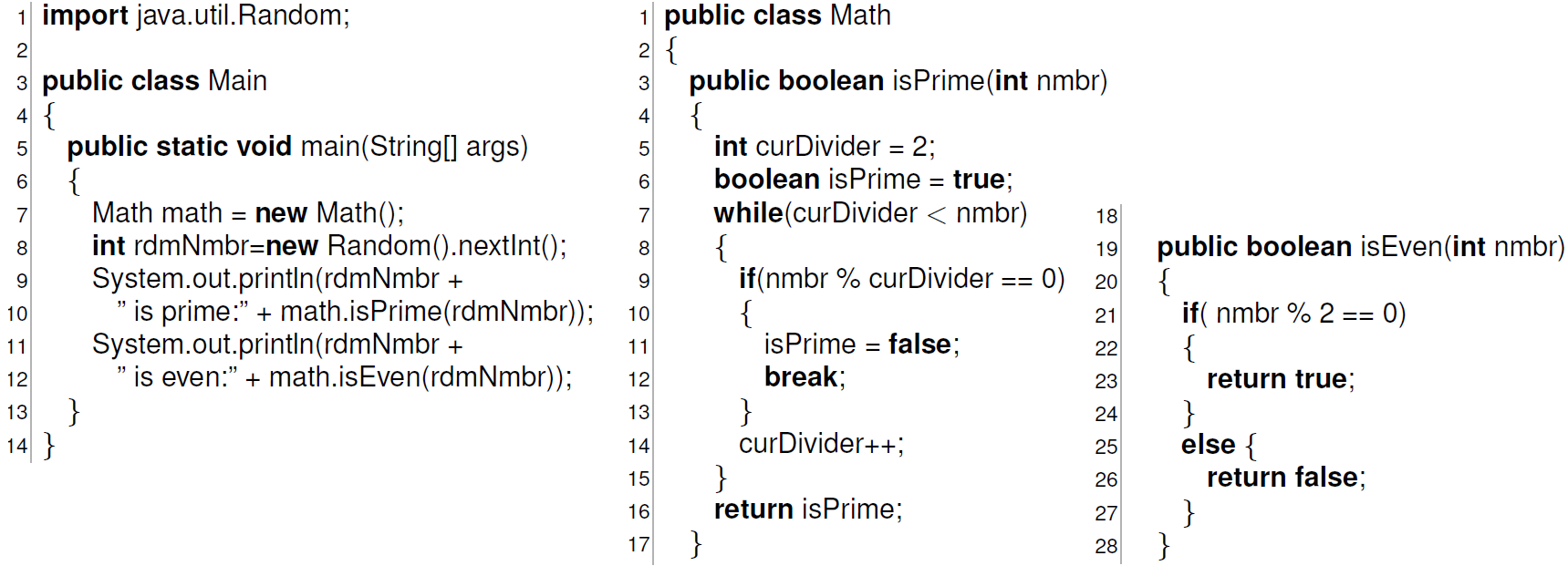}
\caption{Example Java source code}
\label{figExampleCode}
\end{figure}

We use a running example to demonstrate the framework. Figure \ref{figExampleCode} shows the source code of the example Java program. Here, \lstinline{Main.main} generates a random integer, then calculates first if this integer is prime and afterwards if it is even by using methods of the \lstinline{Math} class. We consider this a representative example, since it includes typical imperative object-oriented language structures such as loops, branching points and method invocations.

\begin{figure}
\centering
\includegraphics[width=\textwidth]{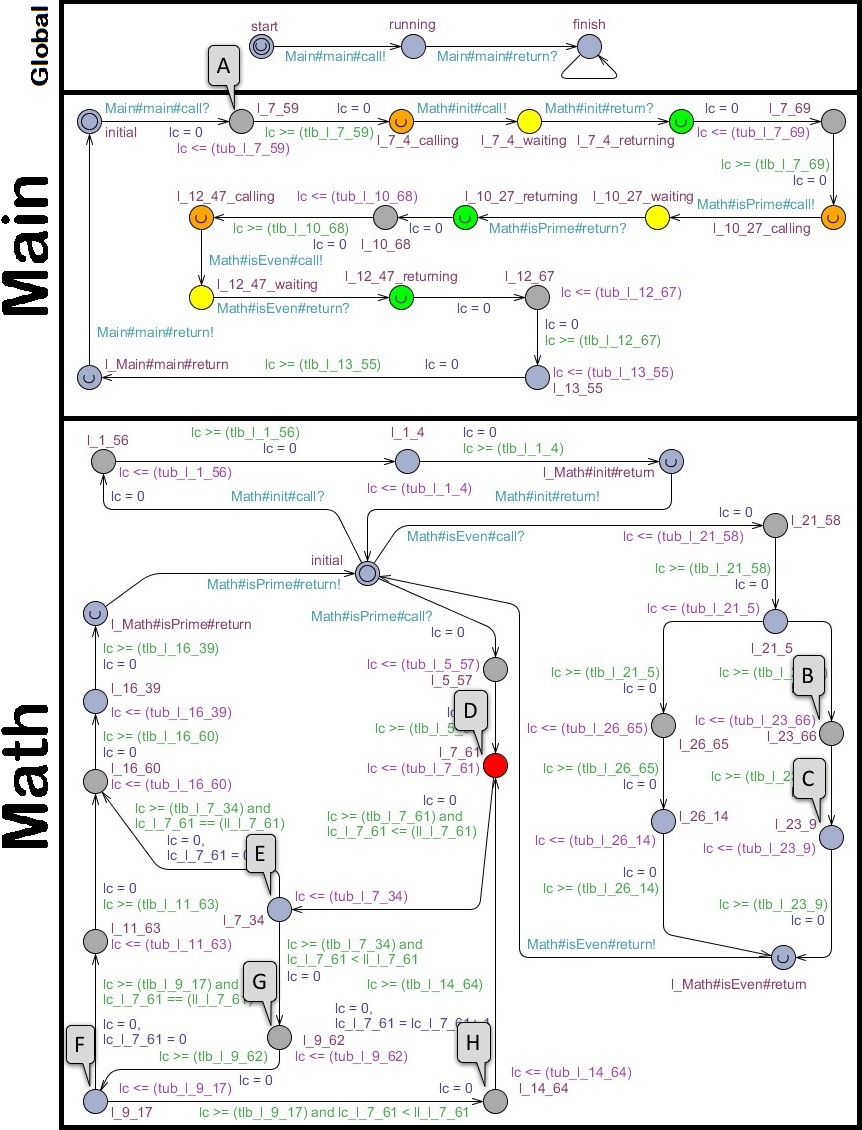}
\caption{Generated timed-automata model from Figure \ref{figExampleCode}} 
\label{figExampleTom}
\end{figure}

The timed-automata model derived from this program is shown in Figure~\ref{figExampleTom}. The first template keeps the global clock, calls the main method of the program and goes to location \lstinline{finish} when the call returns. The second and third templates correspond to the \lstinline{Main} and \lstinline{Math} classes, respectively.
For instance, location \snip{l_7_59} in the second template (see label \snip{A}) corresponds to the bytecode instruction compiled from line~7 of the class \snip{Main} and represents an instruction allocating memory for the object which is constructed at this line. The time spent on the location \snip{l_7_59} is limited between \snip{tlb_l_7_59} and \snip{tub_l_7_59} time units, which are specified by the outgoing edge guard and the location invariant, respectively. The method call for the \snip{Math} object construction is mapped to three locations \snip{l_7_4_calling}, \snip{l_7_4_waiting} and \snip{l_7_4_returning}. The call and return of this method are transformed as a pair of synchronization actions, which are expressed as \snip{Math\hash init\hash call!} and \snip[keywords={}]{Math\hash init\hash return?}, on the corresponding edges. These synchronization actions pass the control flow to the \snip{Math} template for the method execution, then take the control back as soon as the method finishes and returns. 

\medskip\noindent For this example, suppose that we want to check if the main method finishes its execution within \snip{x} time units. This corresponds to checking whether the timed-automata model always reaches \snip{finish} in less than \snip{x} time units, and is expressed by the \uppaal query

\begin{quote}
\snip{A[] main.finish and globalClock <= x}
\end{quote}
This query means that for all paths (expressed through \snip{A}) that begin in the \snip{start} location and terminate at the \snip{finish} location, the condition that \snip{globalClock} is less than or equal to \snip{x} must hold (expressed through \snip{globalClock <= x}).

Suppose that the time spent at each location is 1, the loop limit is 5 and \snip{x} is set to 20. When \uppaal evaluates the query, it reports that the property represented by the query is not satisfied. This means that there is an execution path taking more than \snip{x} time units. In this case, the reported trace visits all the locations in the timed-automata model except the ones with the labels \snip{B} and \snip{C}. This corresponds to an execution of the program where the \snip{else} case of the method \snip{isEven} is executed (lines 25--27).

%% file: actions.tex
\section{Consistency}\label{frameworkActions}\label{sectionConsistency}

In this section, we explain each of the actions shown in Figure \ref{figTifFlow} in more detail. By applying these actions, one can easily re-create timed-automata models for analysis after any changes occur in the code. This helps to have models that are consistent with the software at the time of analysis. 

\paragraph{Action 1: Java Source Code to Bytecode Control-Flow Models.}

To implement Action 1, we have developed JBCPP, which we publish as an Eclipse plug-in. JBCPP generates models from Java bytecode that include an explicit representation of control flow. The models conform to JBCMM (Java Bytecode Metamodel)\footnote{While the term \emph{JBCMM} refers to the metamodel itself, we use \emph{JBCMM model} and \emph{bytecode control-flow model} terms interchangeably to refer to an instance of JBCMM.}, a dedicated metamodel for Java bytecode. The root of a JBCMM model is called \lstinline{Project} and it contains models of all classes whereby the main class is distinguished. A class model in turn contains models for its methods which contain models for all bytecode instructions. The control flow between instructions is modeled through control flow edges.

\paragraph{Action 2: Enriching  Bytecode Control-Flow Models.}

The model resulting from Action 1 faithfully reflects the bytecode. This means, on one hand, it covers all the information needed to execute the program; on the other hand, some information is only presented implicitly. The purpose of Action 2 is therefore twofold. First, details irrelevant for the analysis should be abstracted away. Second, additional information, like loop limits or timing information, should be injected.

\paragraph{Action 3: Bytecode Control-Flow Models to UPPAAL Models.}

The core part of the framework is the transformation definition from JBCMM models to \uppaal timed-automata models. This corresponds to Action 3 in Figure \ref{figTifFlow}. The \uppaal models conform to a metamodel developed by the Software Engineering Group at the University of Paderborn \cite{paderbornRef}. This metamodel consists of all the elements and their relationships of any timed-automata model definable using the \uppaal tool. The metamodel contains the conceptual elements such as locations, edges and clocks; it also contains syntax graphs for the C-like expressions supported by the \uppaal model checker.

\paragraph{Action 4: Generating UPPAAL Files.}

The model conforming to the \uppaal metamodel itself is not directly processable by the \uppaal model checker. To make the model usable by the tool, we have implemented a model-to-text transformation, which takes a \uppaal model as the input and transforms it into an XML file compatible with the \uppaal XML format. The transformation corresponds to Action 4 in Figure \ref{figTifFlow}.

%% file: extensibility.tex
\section{Extensibility}\label{sectionExtensibility}

One can extend the framework in various ways depending on the analysis needs using the following mechanisms: introducing new models with related metamodels or new transformations; extending existing transformations or metamodels; and changing the transformation application order accordingly. In this section, we present several extensions as examples. 

\subsection{Loop Augmentation}\label{sectionLoopAug}

When starting with a piece of source code with loops, the generated timed-automata model will contain cycles. If the number of repetitions would not be limited in the \uppaal model, then the model checker can create unbounded execution paths. In particular this is true if timing is considered in the analysis, as the global clock value would increment infinitely. As a consequence, queries over all execution paths, such as worst case execution time (WCET), will generally not give any meaningful results. To compensate this, we apply two steps:

\emph{Loop Detection:} The loop detection takes a JBCMM model as input, detects the loops in it and generates an extended JBCMM model as output which additionally represents loop-related information such as instructions in the loop and loop limit. We use a dominator analysis \cite{dominatorAnalysisReference} for loop detection.

The framework provides an interface through the extension of JBCMM to allow its user to insert loop limits. The loop limits to be inserted can be obtained using manual annotations \cite{bernat} or automated analysis techniques \cite{lokuciejewski2009}. Currently, we use a default value as the loop limit for all the loops. 

\emph{Insertion of Loop Information:} Once the loops are detected and the extended JBCMM model is generated, loop-related information is inserted into the timed-automata model during Action 3. For this purpose, we have implemented a transformation module, which is called \emph{Loop Information Insertion}. This module extends the transformation from bytecode control-flow models to \uppaal timed-automata models (Action~3) by simply adding one additional transformation rule for loops and reusing all the other rules in the transformation. As an example how this module works, consider Figure~\ref{figExampleTom}. The location \lstinline{l_7_61} (label \lstinline{D}) is the head of the loop. The counter and the repetition limit are named after the loop head as \lstinline{lc_l_7_61} and \lstinline{ll_l_7_61}, respectively. The locations \lstinline{l_7_34} (label \lstinline{E}) and \lstinline{l_9_17} (label \lstinline{F}) are the  exiting points from the loop. The locations \lstinline{l_9_62} (label \lstinline{G}) and \lstinline{l_14_64} (label \lstinline{H}) are the remaining loop nodes. On the back edge, which is from the location \lstinline{l_14_64} (label \lstinline{H}) to the head location, the loop counter is incremented. The exiting edges are guarded by the condition \lstinline{lc_l_7_61==ll_l_7_61}, which checks if the loop counter already reached the limit; the continuing edges are guarded by the condition \lstinline{lc_l_7_61 <ll_l_7_61}, which checks if the loop counter is still below the limit. The loop counter is reset on the exiting edges.

\subsection{Recursion Handling}\label{recursionSection}	
A template instance in the \uppaal model represents a stack frame in a Java program execution. The number of template instances for each template needs to be known before using the model checker for the timing analysis since \uppaal does not allow to create new instances on the fly. If we start with fewer instances than the possible call stack size, then we will end up in a deadlock state. Therefore, our framework currently handles recursive calls by removing direct recursion (calls to the method containing the invocation) and reporting other forms of recursion.

For removing the direct recursive calls, we have implemented a transformation called \emph{Recursion Removal}. The \emph{Recursion Removal} transformation takes a JBCMM model and outputs a new JBCMM model in which the direct recursive call instructions are replaced by some dummy instructions.

For reporting other forms of recursion, we have implemented a transformation to derive the call graph of the JBCMM model. A cycle in such a graph shows the existence of a recursive call. The transformation generates a call graph from a JBCMM model and reports any recursive call structures by analyzing this call graph. 

\subsection{Timing Augmentation}

The \emph{Timing Augmentation} transformation takes a JBCMM model as input and enhances it with the timing information, i.e., the minimum and maximum time spent for each instruction's execution. Currently, we use default values for the timing information. The framework provides an interface through slots in JBCMM models for the timing values to allow its user to insert timing information. The user can acquire timing information to be inserted by using various techniques such as profiling \cite{puschner2000guest} or JVM Timing Models \cite{lambert2008platform, hu2003deriving}.

%% file: experiments.tex
\section{Scalability}\label{experiment}\label{sectionScalability}

In this section, we will present evidence for the scalability of the framework. For our purpose, the framework should be able to cope with realistic software sizes. To assess whether this is the case, we have chosen three real-life open source Java programs of different sizes as input.

Table~\ref{tableModelElementCounts} shows the characteristics of their derived JBCMM models. The columns A through E show the counts of corresponding elements in the model. Column F shows the count of the method call instructions whose invokable method implementations are included in the model. Column G shows the total number of possible method implementations invocable by method calls. Column H shows the count of the return instructions. The size of a program can be determined by the number of model elements that its JBCMM model contains: A+B+C+D+E.

\begin{table}[h]
\centering
\resizebox{1\textwidth}{!}{\begin{minipage}{1.2\textwidth}
\begin{tabularx}{1\textwidth}{Xrrrrrrrrr}
\toprule
&
\multicolumn{1}{c}{Class} &
\multicolumn{1}{c}{Method} &
\multicolumn{1}{c}{Loop} &
\multicolumn{1}{c}{\textbf{}Instruc-} &
\multicolumn{1}{c}{Edge} &
\multicolumn{1}{c}{Method} &
\multicolumn{1}{c}{Method} &
\multicolumn{1}{c}{Return} &
\multicolumn{1}{c}{Total (A+B}\\
&
\multicolumn{1}{c}{(A)} &
\multicolumn{1}{c}{(B)} &
\multicolumn{1}{c}{(C)} &
\multicolumn{1}{c}{tion (D)} &
\multicolumn{1}{c}{(E)} &
\multicolumn{1}{c}{Call (F)} &
\multicolumn{1}{c}{Invoc. (G)} &
\multicolumn{1}{c}{Instr. (H)} &
\multicolumn{1}{c}{+C+D+E)} \\ \midrule

LiveGraph & $131$ & $350$ & $33$ & $11795$ & $11,740$ & $665$ & $687$ & $440$ & $24,049$ \\
Groove Generator & $930$ & $5,392$ & $756$ & $99,738$ & $98,634$ & $9,790$ & $12,718$ & $7,114$ & $205,450$ \\
Groove Simulator & $1,482$ & $9,232$ & $1,454$ & $203,030$ & $203,071$ & $20,198$ & $252,72$ & $12,101$ & $418,269$ \\
Weka & $1,041$ & $83,22$ & $4,072$ & $367,774$ & $374,854$ & $30,124$ & $108,570$ & $10,820$ & $756,063$ \\

\bottomrule
\end{tabularx}
\caption{Characteristics of the JBCMM Models of the Example Programs}
\label{tableModelElementCounts}
\end{minipage}}
\end{table}

\emph{LiveGraph} is a real-time graph and chart plotter to represent large amounts of data \cite{liveGraphWebsiteRef}. \emph{Groove} is a tool for modeling and analyzing object-oriented systems through graphs and graph transformations \cite{grooveWebsiteRef}. We have examined the \emph{Simulator} and \emph{Generator} components of Groove. \emph{Weka} offers a large collection of machine learning algorithms with pre-processing of the data and visualization of the results \cite{wekaWebsiteRef}. Although the class and method counts are close to the Groove components, the total model size is around 1.8 times as much as the Groove Simulator due to the large instruction and edge counts.


\subsection{Scalability of the Framework Actions}\label{scalFASection}

We define the scalability of the framework actions as the ability to get acceptable time performance measurements with the increase in the input size. 

\emph{Prediction on outcome: }Although there is an expectancy to observe a linear dependency between the timing performance and the derived model sizes for Action 1, the performance can depend on various factors such as the number of processed classes on the classpath (which is different from the number of classes included in the model) in the example Java programs. For Action 2, we have implemented a practically fast implementation of the dominator analysis algorithm of the complexity $O(D^2)$ for loop detection, so we expect the timing performance to have at worst a quadratic dependency with respect to the model size (but in practice, it can run faster) \cite{dominatorAnalysisReference}. We expect to get a linear association between the timing performance and the model sizes for the actions 3 and 4 since these model transformations are direct mappings of input elements to output elements and do not have any special algorithmic computations used, unlike Action 2.

\emph{Outcome: }We have applied the actions 1 through 4 to obtain the \uppaal textual model of each program.  The timing results of these experiments are presented in Table~\ref{tableExperimentResults}. Each action has been repeated 10 times for each program, the table shows the averages. The experiments have been carried out using an Intel i7-3520M 2.90 GHz CPU with 4 cores and 16 GB RAM.

\begin{table}[h]
\centering
	\resizebox{1\textwidth}{!}{\begin{minipage}{1.2\textwidth}
\begin{tabularx}{\textwidth}{Xrrrrrr}
\toprule

&
\multicolumn{1}{c}{Model Size} &
\multicolumn{1}{c}{Automatic Model} &
\multicolumn{1}{c}{Loop} &
\multicolumn{1}{c}{JBCMM to UPPAAL} &
\multicolumn{1}{c}{Model-to-Text} &
\multicolumn{1}{c}{Total} \\

&
\multicolumn{1}{c}{} &
\multicolumn{1}{c}{Derivation} &
\multicolumn{1}{c}{Augmentation} &
\multicolumn{1}{c}{Transformation} &
\multicolumn{1}{c}{Transformation} &
\multicolumn{1}{c}{Time} \\

&
\multicolumn{1}{c}{} &
\multicolumn{1}{c}{(Action 1)} &
\multicolumn{1}{c}{(Action 2)} &
\multicolumn{1}{c}{(Action 3)} &
\multicolumn{1}{c}{(Action 4)} &
\multicolumn{1}{c}{(sec)} \\
\midrule

LiveGraph & $24,049$ & $18$ & $51$ & $12$ & $35$ & $117$ \\
Groove Generator & $205,450$ & $1,414$ & $86$ & $194$ & $364$ & $2,058$ \\
Groove Simulator & $418,269$ & $1,480$ & $300$ & $538$ & $977$ & $3,295$ \\
Weka & $756,063$ & $764$ & $803$ & $1,069$ & $2,402$ & $5,037$ \\
\bottomrule
\end{tabularx}
\caption{Experiment Results (in seconds, averaged over 10 runs)}
\label{tableExperimentResults}
	\end{minipage}}
\end{table}

\emph{Evaluation of the outcome: }For Action 1, the results show no particular relationship with the derived model sizes. We tried to find a correlation with various possible factors related to this Action, but currently we cannot say what is the determining factor. Nevertheless, the timing performance is still acceptable for large projects like Weka.

Although the algorithmic complexity of the dominator analysis algorithm that we have used is quadratic with respect to the number of instructions in the input models, the loop augmentation transformation with the practically fast implementation of the dominator analysis algorithm still runs only around 13 minutes for the largest of our input programs. The last two columns of Table~\ref{tableExperimentResults} show the timing performance of the actions 3 and 4. The figures support the hypothesis that the performance of the model-to-model and model-to-text transformations are linear with respect to the input size of the models. The experiments show that the timing performance of the framework scales well with respect to the varying code sizes.

\subsection{Scalability of Model Checking of the Generated UPPAAL Models}\label{scalMCSection}

Although model checking itself is outside the scope of this paper, let us discuss some ways to improve scalability of model checking of the generated models.

It is a major challenge to adjust the correct abstraction level for models to avoid the state-space explosion problem in model checking. The more detailed the models are, the more accurate the results one can get. However, increasing the detail level of models can cause intractable state-space sizes. Raising the abstraction level, in our approach, can be achieved by extending our framework with new transformations. We have implemented such an extension that abstracts away some details by grouping nodes.

Furthermore, \uppaal provides some generic mechanisms to reduce the state-space size or to optimize the state-space generation/exploration by removing redundancy. We have defined our strategy for transforming JBCMM models to \uppaal timed automate models such that these mechanisms are applicable. Both optimizations are detailed below, followed by a discussion of their effects.

\emph{Node Grouping:} This extension allows to reduce the state-space size by decreasing the number of locations in the timed-automata model, by replacing sequences of bytecode instructions (connected with control-flow edges) with a \emph{group} instruction that accumulates the timing characteristics of its instructions. Method calls and branches are not grouped since they affect the execution flow.

\emph{Symmetry Reduction:} When a template instance $T_A$ has to synchronize with an instance of template $T_B$, it needs to choose with which instance it wants to synchronize. \uppaal generates the same state-space for any choice of instances of template $T_B$ if all instances of the same template are identical. For such a case, \uppaal can be guided to not to generate redundant states, which is called symmetry reduction optimization. For this reason, we define templates such that no extra states are introduced for identical cases, enabling the symmetry reduction optimization of \uppaal in our default transformation definition from JBCMM models to \uppaal models.

\paragraph{Experiment setup and outcome.}

We have used the example Java program given in Section \ref{frameworkActions} to test how much Node Grouping and Symmetry Reduction help to reduce the state space. To check the maximum possible size of the state-space of the \uppaal model, we have chosen a query that checks whether synchronization points in \uppaal can be blocking each other. As we only use synchronization actions to represent method invocation, they are properly nested in our case and deadlocks cannot appear. Therefore, such a query will lead to exploring the whole state space of the model. For this example case, we have achieved a reduction of around 80\% in the size of the generated state-space when both optimizations have been applied. The details of this experiment can be found in the technical report \cite{eemcs26622}.

\section{Expressiveness}\label{exampleAnalysisExp}\label{sectionExpressiveness}

We have conducted an example analysis of LiveGraph to show how we can use the models of real-life programs generated by our framework for analysis. LiveGraph stores the data to be plotted in the \snip{DataCache} class. The changes in the instances of this class trigger firing of an event via the \snip{fireEvent} method of the same class, which notifies observers. 

As an example analysis, we want to check if resetting of data labels, \snip{DataCache.resetLabel}, eventually triggers firing of an event. This kind of check is common in program analyses. The following query is the formulation of this property (note that we simplified the names for readability): 

\begin{minipage}{\linewidth}
\begin{lstlisting}[keywords={exists}]
(exists (id:DataCacheTemplateId) DataCacheTemplate(id).l_resetLabelF) -->
(exists (id:DataCacheTemplateId) DataCacheTemplate(id).l_fireEventF)
\end{lstlisting}
\end{minipage}

The location \snip{l_resetLabelF} is the first to be visited when \snip{resetLabel}  is called. Similarly, \snip{l_fireEventF} is the first to be visited when \snip{fireEvent} is called. Both locations are in the same template, DataCacheTemplate, which corresponds to the \snip{DataCache} class. The \snip{-->} sign specifies the \emph{eventually leads to} statement. The \lstinline[keywords={exists}]{exists} statements are used to make sure that these locations can be reached by any instance of the template. The query asks if the visit of the location \snip{l_resetLabelF} eventually leads to the visit of the location \snip{l_fireEventF}, which corresponds to what we want to check.

This example shows that the models generated by the framework provide a practical way for conducting desired analyses through defining small and intuitive queries in a simple way.


%% file: concl.tex
\section{Conclusion}\label{sectionConclusion}

In this paper, we have presented a framework that derives models from Java programs in an automatic way for analysis. We have shown that the framework provides the following features:

\begin{enumerate}[leftmargin=*]
\item \emph{Consistency}: Since the model derivation process consists of multiple actions that are fully automated, it is easy to re-create models after changes occur in the code. Therefore, there is no need to separately maintain code and models, and this prevents inconsistency.

\item \emph{Extensibility}: Due to adoption of MDE techniques, users can adapt and extend the framework conveniently. Metamodels as well as transformations can easily be extended. We have demonstrated this by developing four different transformations to enrich the models derived by our framework.

\item \emph{Scalability}: We have derived models with our framework from large Java programs with up to 1,482 classes and up to 367,774 bytecode instructions. The largest model that we derived in our study contained 756,063 nodes. The longest time our framework took to derive a model was 84 minutes, which is still sufficient to re-create models on a nightly basis. For a smaller project of 131 classes and 11,795 instructions, model derivation took only 2 minutes.

\item \emph{Expressiveness}: We have described a translation scheme to create \uppaal timed-automata models from the ECore-based model that we derive. The timed-automata model faithfully reflects the execution semantics of the Java program, in particular for method invocation and branching instructions. The translation scheme is also provided as model transformation as part of our framework.
\end{enumerate}

In addition to these features, the automation of the model derivation process reduces errors that can be caused by manual processes; and eliminates variations in models caused by subjective decisions.

With our Java bytecode metamodel, we have presented a generic way of modeling compiled Java code. By using EMF-technology, we facilitate a high degree of interoperability between multiple static Java bytecode analyses.

\paragraph{Future work.} One direction for future work is to include support for data-flow analysis to increase the precision of derived information such as loop limits or timing values. 

Another direction is to include stochastic information to allow users of the framework to do probabilistic model checking with the derived timed-automata models. 

Yet another direction is to provide support for concurrent Java programs. The instances of templates in UPPAAL models are time automata (processes) running in parallel, by definition. In the future, we plan to adapt the model transformation in Action 3 (from bytecode models to UPPAAL models) to treat concurrency-related constructs in bytecode models, such as new tread creations or synchronization directives, specifically to analyze concurrent programs by benefiting from the parallelism of time automata in UPPAAL models. 

Finally, we plan to provide a convenient front-end for using \uppaal together with models created by our framework; this will allow to perform analyses and interpret results in terms of the source code rather than locations in the timed-automata model.